\documentstyle [12 pt] {article}

\title{QUANTUM SUPERPOSITION OF A MIRROR AND RELATIVE DECOHERENCE
(AS SPONTANEOUS SUPERPOSITION BREAKING)}

\author{Vladan Pankovi\'c,Milan Predojevi\'c,Miodrag Krmar$^{\ddag}$\\
Institute of Physics,Faculty of Natural Sciences and
Mathematics,\\21 000 Novi Sad,Trg Dositeja Obradovica 4.\\Serbia
and Montenegro, mpgalant@ptt.yu\\
$^{\ddag}$ California State University Dominguez Hills\\
1000 E Victoria Street, Carson, California, mkrmar@csudh.edu}
\date {}

\begin {document}
\baselineskip 20 pt

\maketitle


\begin{abstract}

In this work recently suggested (by Marshall, Simon, Penrose and
Bo\-uw\-me\-es\-ter) experiment of the quantum superposition of a
(qua\-si)\-mac\-roscopic mirror (an oscillating part of a
Michelson interferometer) interacting with a single photon is
consequently interpreted by relative decoherence. Namely, it is
shown that relative decoherence (based on the spontaneous unitary
symmetry (superposition) breaking (effective hiding) ) on the
photon caused by mirror is sufficient to model real measurement
with photon as a measured quantum object and mirror as a
measurement device.
\end{abstract}

\newpage

\section{Introduction}

Recently Marshall, Simon, Penrose and Bouwmeester $\cite{jed}$
suggested a very interesting and  important experiment. Theirs
primary aim is an experimental realization of an analogy of
Schrödinger cat paradox $\cite{dva}$.

Precisely it was creation of a realistic, i.e. observable, quantum
superposition on a mesoscopic, even (quasi)macroscopic system
(with $~ 10^{14}$ atoms and linear dimensions of $10\mu m$). It is
a tiny mirror (a oscillating part of a Michelson interferometer
with high-fines cavities) interacting with a single photon. More
precisely speaking there is a quantum dynamical interaction
between single photon (being initially in a quantum superposition)
and mirror (being initially described by a wave packet)  that do
commonly a quantum supersystem photon+mirror. In conditions of the
extremely low temperature (less than 2 mK, which  eliminate any
environmental thermical absolute decoherence) given dynamical
interaction periodically, i.e. alternately correlates (entangles)
photon and mirror into photon+mirror  or decorrelates
(disentagles) photon+mirror into photon and mirror. Dynamical
restitution of correlation breaks effectively previous quantum
superposition of the photon and does that  the mirror be described
by a quantum superposition of the wave packets (more precisely
speaking it is described by a second kind mixture $\cite{tri}$
that includes given quantum superposition). But dynamical
decorrelation breaks effectively previous quantum superposition of
the mirror and reverts a quantum superposition of the photon which
can be simply  experimentally tested (by detection of the photon
interference effects). Quantum superposition of the photon would
not be reverted in case that by correlation any absolute
decoherence on photon+mirror, as well as on mirror itself appears.
For this reason revival of the quantum superposition of the photon
can be considered as an immediate and retrospective proof that
mirror previously, i.e. during correlation has been in a quantum
superposition.

Second aim of the authors, i.e. : "one long-term motivation for
this kind of experiment is the search for unconventional
decoherence processes" $\cite{jed}$. In other words authors
considerate implicitly different possibilities (without standard
quantum mechanical formalism $\cite{tri}$ - $\cite{pet}$ and its
usual interpretation  $\cite{ses}$ , $\cite{sed}$ for an {\it
absolute decoherence (collapse) } on photon+mirror.

In this work suggested experiment of the quantum superposition of
given mirror will be consequently analyzed in the conventional
way, i.e. from view point of the standard quantum mechanical
formalism including its usual interpretation.

Namely an important theorem will be proved that any exact quantum
superposition of the wave packets (that does not represent any
wave packet) turns effectively approximately, i.e. spontaneously
and probabilistically (spontaneous unitary symmetry
(superposition) breaking (effective hiding)) in some of wave
packets from superposition. It admits that a {\it relative and
effective decoherence (collapse)} be consequently defined as a
special case of the general formalism of the spontaneous symmetry
breaking (applicable in many different domains of the physics, eg.
quantum field theory, quantum theory of ferro-magnetism, classical
mechanics of the deformable bodies, etc.)
$\cite{osa}$,$\cite{dev}$.

Such relative and effective decoherence represents, of course, a
consequent formalization of Bohr principle of the {\it relative
boundary} (even {\it complementarity}) between measured quantum
object and measurement device   $\cite{ses}$, $\cite{sed}$. Bohr
said : "Especially, the singular position of the measuring
instruments in the account of the quantum phenomena, just
discussed, appears closely analogous to well-known necessity in
relativity theory of upholding an ordinary description of all
measurement process, including a sharp distinction between space
and time coordinates, although the very essence of this theory is
the establishment of new physical law, in the comprehension of
which we must renounce the customary separation of space and time
ideas. The dependence of the reference system, in relativity
theory, of all readings of scales and clocks may even be compared
with the essentially uncontrolable exchange of the momentum or
energy between the objects of measurements and all instruments
defining the space-time system of the reference, which in quantum
theory confronts us with the situation characterized by notion of
complementarity. In  fact this new feature of natural philosophy
means a radical revision of our attitude as regards physical
reality, which may be paralleled with the fundamental modification
of all ideas regarding the absolute character of physical
phenomena, brought about by the general theory of relativity."
$\cite{ses}$

Finally, it will be proved that in given Marshall {\it et al.}
experiment of a quantum superposition of a mirror all conditions
for relative decoherence on the correlated photon+mirror are
satisfied so that in this case photon can be consequently treated
as a measured quantum object and mirror as a measurement device.
So, mirror can be exactly treated as a (quasi)macroscopic object
in a quantum superposition (Schrödinger cat), but complementary,
i.e. effectively approximately it can be treated as a
(quasi)microscopic measurement device.

All this indicates, quite generally, that {\it any absolute
decoherence is not necessary in any measurement process}. Thus, it
will be suggested that, like remarkable Michelson experiment that
affirmed Einstein relativistic statement on the absence of the
absolute space, Marshall et al. experiment on the quantum
superposition of a mirror (really included in a Michelson
interferometer)  would to affirm Bohr statement on the absence of
the absolute decoherence (collapse) and on the completeness of the
standard quantum mechanical formalism.

\section{Relative  decoherence as a spontaneous unitary symmetry (superposition) breaking}

Within standard quantum mechanical formalism $\cite{tri}$ -
$\cite{pet}$ dynamics of a quantum system is described by
Schrödinger equation
\begin{equation}
\hat{H} \Psi= i\hbar \frac{\partial \Psi}{\partial t}
\end{equation}
where $\hat{H}$ represents corresponding Hamiltonian observable,
$\Psi$ - quantum state of the unit norm from Hilbert space $\cal
H$, $t$ - time moment and $\hbar$ reduced Planck constant. Formal
solution of (1) can be presented by expression
\begin {equation}
\Psi (t)={\bf U}(t) \Psi_{0}
\end {equation}
where ${\bf U}(t)$ represents corresponding unitary evolution
operator and $\Psi_{0}$ initial quantum state. Obviously, there is
deterministic, i.e. one-to-one correspondence between initial and
final quantum state.

Let ${\cal B}=\{\Psi_{n}, \forall n\} $ be an arbitrary time
independent basis in $\cal H$  and let ${\bf U}(t) {\cal B}=\{{\bf
U}(t) \Psi_n, \forall n\}$ represents basis in $\cal H$ that
represents the dynamical evolution of $\cal B$. Then it follows
\begin {equation}
\Psi  (t) = \sum_{n}(\Psi_{n}(t),\Psi (t)) \Psi_{n}(t) = \sum_{n}
(\Psi_{n},\Psi_{0}) \Psi_{n}(t)
\end {equation}
where
\begin {equation}
\sum_{n} |(\Psi_{n}(t),\Psi
(t))|^{2}=\sum_{n}|(\Psi_{n},\Psi_{0})|^{2}=1
\end {equation}
In  sense of (3) and (4) quantum mechanical dynamical evolution
conserves superposition and unit norm of the quantum state during
time and it does not prefer any basis in $\cal H$. It represents a
fundamental symmetry of quantum mechanical dynamics, called {\it
unitary symmetry}, which practically means that all referential
frames, i.e. all bases in Hilbert space can be treated as
completely same right. In this way unitary symmetry represents a
{\it very important}  symmetry that connects the kinematical
aspects (characteristics of Hilbert space) and dynamical aspects
of the quantum mechanics, like Galilei symmetry within classical
mechanics or Lorentz symmetry within special theory of relativity.

But there is following, seemingly unambiguous, experimental fact.
By a measurement of some observable with  eigen basis $\cal B$
realized on the measured quantum object O, in quantum state $\Psi
(t)$ given state turns {\it exactly}  in some quantum state
$\Psi_{n}$ from $\cal B$ with probability
$w_{n}(t)=|(\Psi_{n},\Psi (t))|^{2}$ for arbitrary $n$. Duration
of the measurement is relatively very short, so that during this
measurement effects of the dynamical evolution of isolated O can
be neglected. In this way given measurement {\it breaks exactly
and absolutely} unitary symmetry (does {\it absolute exact
collapse}), since even if it conserves unit norm of the quantum
state it breaks superposition. For this reason within standard
quantum mechanical formalism measurement cannot be consistently
presented by any exact quantum mechanical dynamical evolution on
O.

Moreover, von Neumann proved $\cite{cet}$ that within standard
quantum mechanical formalism measurement cannot be presented by
any exact quantum mechanical dynamical evolution even on the
quantum supersystem O+M, consisting from dynamically interacting
quantum subsystems, O and measurement device M. This von Neumann
proof is conceptually very simple.

Namely, let O before measurement be in a quantum state
\begin {equation}
\Psi^{O}=\sum_{n} c_{n} \Psi^{O}_{n}
\end {equation}
Here ${\cal B}_{O}=\{\Psi^{O}_{n} , \forall n\}$ represents eigen
basis of the measured observable in Hilbert space ${\cal H}_{O}$
of O, while $c_{n}$ for $\forall n$ represent superposition
coefficients that satisfy normalization condition
\begin {equation}
\sum_{n} |c_{n}|^{2}=1
\end {equation}
Let before measurement M be in a quantum state $\Psi^{M}_{0}$ from
${\cal B}_{M}=\{\Psi^{M}_{n} , \forall n\}$ that represents eigen
basis of the so-called pointer observable in Hilbert space ${\cal
H}_{M}$ of M. Then before measurement O+M is described by
noncorrelated quantum state
\begin {equation}
\Psi^{O} \otimes \Psi^{M} = \sum_{n} c_{n} \Psi^{O}_{n}\otimes
\Psi^{M}_{0}
\end {equation}
from ${\cal H}_{O} \otimes {\cal H}_{M}$ ,where $\otimes$
represents tensorial product.

Usual empirical facts that characterize measurement need that
unitary evolution operator ${\bf U}_{O+M}$ corresponding to
measurement be determined in such way that it restitutes
one-to-one correspondence between $\cal B_{O}$ and $\cal B_{M}$
without any changing of the superposition coefficients in (7). It
simply mean that after measurement treated as a quantum mechanical
dynamical evolution on O+M this O+M  must be finally in the
following correlated quantum state
\begin {equation}
\Psi^{O+M}=\sum_{n} c_{n} \Psi^{O}_{n} \otimes \Psi^{M}_{n}
\end {equation}
from ${\cal H}_{O} \otimes {\cal H}_{M}$.

However within standard quantum mechanical formalism $\Psi^{O+M}$
(as the final result of the quantum mechanical dynamical evolution
on O+M)  is different from any  pure or mixed quantum state of O+M
that can correspond to final empirical  measurement results.
Precisely within standard quantum mechanical formalism
$\Psi^{O+M}$ is different from any   $\Psi^{O}_{n} \otimes
\Psi^{M}_{n}$ or from mixture of such quantum states with
corresponding statistical weights $w_{n}=|c_{n}|^{2}$ for $\forall
n $.

In this way, according to von Neumann proof, deterministic
dynamical evolution and probabilistic measurement represent two
exact but quite independent and different ways of the change of
the quantum state within standard quantum mechanical formalism.
Origin of the quantum mechanical dynamical evolution is more or
less clear, while origin ("place and time") of the measurement,
precisely decoherence (collapse) is completely unclear and it must
be ad hoc postulated (von Neumann projection postulate), like
absolute space in classical mechanics. In any way it represents a
serious open problem in quantum mechanics foundation $\cite{des}$.

There are two opposite attempts of the solution of given problem.
First one accepts unitary symmetry (superposition) breaking
(decoherence, collapse) by measurement as {\it an absolute exact}
phenomenon. Simultaneously it considers that standard quantum
mechanical formalism is incomplete (even contradictory) and that
it must be extended by some unconventional but complete theory.
But, as it is well-known $\cite {Jed}$ - $\cite {Tri}$, such
solution is very nonplausible since it predicts basic
contradiction between quantum mechanics and theory of relativity
and forbids building of any consistent quantum field theory.

Second one attempt of the solution of measurement or decoherence
(collapse)  problem represents Bohr relative boundary between O
and M or relative decoherence (collapse) principle (this principle
represents general form of the Bohr complementarity principle)
$\cite {ses}$, $\cite {sed}$. It supposes that quantum mechanical
dynamical evolution represents {\it unique completely exact} form
of the quantum state change which implies that {\it unitary
symmetry stands always exactly conserved}. Also, it means that
final result of the completely exact interaction between O and M
is given by $\Psi^{O+M}$ (8), so that {\it absolute boundary}
between O and M or {\it absolute decoherence ( collapse) does not
exist at all }.

But introduced supposition needs that measurement be defined in an
especial way but within standard quantum mechanical formalism
exclusively. It seems that Bohr suggested that exact decoherence
(collapse) on the quantum mechanically exactly described O is {\it
not absolutely exact } but only {\it relatively and effectively
exact}. Namely, it appears only {\it in relation} to approximately
quantum mechanically, i.e. "classical mechanically" described M.
In this sense given  "classical" description of M generates
effective and relative boundary between quantum described O and
"classically" described M. More precisely speaking $\cite {Cet}$,
$\cite {Pet}$ ${\cal B}_{M}$ must be an {\it especially chosen
basis of weakly interfering wave packets} (which means that the
distance between centers of any two wave packets must be greater
than any wave packet width), while ${\cal B}_{O}$ is exactly
uniquely correlated ${\cal B}_{M}$ by ${\bf U}_{O+M}$. Obviously
given exact unique quantum  dynamical correlation between ${\cal
B}_{O}$ and ${\cal B}_{M}$ forbids any simultaneous exact
unambiguous quantum dynamical correlation between any basis
incompatible with ${\cal B}_{O}$ in ${\cal H}_{O}$ and ${\cal
B}_{M}$. In other words it means that mutually noncommutative
observable of O can be measured neither simultaneously nor by the
same measurement device which, obviously, corresponds to
Heisenberg uncertainty relations.

Thus, {\it completely exact}  quantum description of O+M (without
any boundary between O and M) and {\it hybrid} description of O+M
(effectively exact description of O and "classical" description of
M) exist simultaneously but on the discretely different levels of
the analysis accuracy. Any of these two levels can be chosen quite
arbitrary but relative and effective decoherence  exists on the
level of hybrid description only. In this way Bohr principle of
the relative boundary and relative decoherence  yields a solid
basis for understanding of the measurement process within standard
quantum mechanical formalism without any contradiction (in
previously noted sense $\cite{Jed}$ - $\cite{Tri}$) with theory of
relativity or quantum field theory.

However, as it has been pointed out $\cite {des}$, measurement
cannot be completely formalized by presented suppositions on the
relative boundary or relative decohenrece principle only. Namely,
even for ${\cal B}_{M}$ as a basis of weakly interfereing wave
packets $\Psi^{O+M}$ conserves the same form (8) that cannot to
explain appearance of $\Psi^{O}_{n} \otimes \Psi^{M}_{n}$ with
probability $w_{n}$ for arbitrary n in case of an individual
measurement.

Nevertheless, noted problem of the consequent formalization of the
relative boundary or relative decoherence principle can be solved
in following way. It can be observed that a physical state is
dynamically stable if it satisfies corresponding dynamical
equation, i.e. corresponding Hamilton least action principle. In
this sense {\it a quantum state is quantum mechanically always
dynamically  but it is not always classical mechanically
dynamically stable}. Precisely speaking {\it a quantum state is
classical mechanically dynamically stable only in wave packet
approximation while in all other cases it is classical
mechanically dynamically nonstable}. As it is well-known
$\cite{tri}$, $\cite{pet}$ a quantum state $\Psi$ satisfies wave
packet approximation if following is satisfied
\begin{equation}
|(\Psi,\hat {A} \Psi)| \gg ( (\Psi,\hat {A}^{2}\Psi) - (\Psi,\hat
{A} \Psi)^{2})^{\frac{1}{2}}
\end {equation}
i.e. if absolute value of average value of any observable
$\hat{A}$ from orbital Hilbert space in $\Psi$ is  many times
greater that the standard deviation $\hat{A}$ in $\Psi$.

Further it is not hard to see that following is satisfied. {\it
Any quantum state representing a (nontrivial) superposition of
weakly interfering wave packets from some basis of Hilbert space
does not represent any wave packet. It means that any
superposition of the weakly interfering wave packets is quantum
mechanically dynamically stable but that it is classical
mechanically dynamically nonstable}. For this reason given
superposition without any its change on the quantum level of
analysis accuracy turns spontaneously (dynamically nonobservably
and probabilistically), under condition of the conservation of
unit norm of the quantum state, in its arbitrary wave packet on
the "classical mechanical" level of the analysis accuracy. This
spontaneous unitary symmetry (precisely superposition) breaking
(effective hiding), representing an especial case of the general
formalism of the spontaneous symmetry breaking (applicable in
different domains of the physics: quantum field theory, quantum
theory of the ferro-magnetism, classical mechanics of the
deformable bodies, etc.) $\cite {osa}$, $\cite {dev}$, can be
treated as a {\it selfdecoherence (selfcollapse)}. In this way an
important {\it  theorem on the selfdecoherence as spontaneous
unitary symmetry (superposition) breaking (effective hiding)} is
proved.

Such "classical mechanical" selfdecoherence appears obviously on M
and, by means of the previously quantum mechanically dynamically
realized correlation between O and M, it manifests itself
mediately as an {\it effective exact quantum mechanical relative
decoherence(collapse)} on O .

So, it can be concluded that Bohr principle of the relative
boundary or relative decoherence  can be consequently formalized
within standard quantum mechanical formalism by means  of the
spontaneous unitary symmetry (superposition) breaking.

In this way it can be stated that, there are, practiaclly only two
possible general ways for solution of the problem of the quantum
measurement: first one with absolute decoherence and second one
with relative decoherence. Independently from noted theoretical
nonplausibility of the absolute decoherence theories and
theoretical plausibility of the relative decoherence  theories
final decision which of these two general types of theories is
physically acceptable can be obtained only by means of a
convenient experiment (measurement). Such experiment must to
detect exact quantum superposition effects, if they exist, on O+M,
and, without any principal changing (except corresponding
approximation) of the dynamical interaction between O and M,
decoherence effects, if they exist, on O+M. In case of the
positive detection in both cases it can be stated that M realizes
relative decoherence  on O and that experiment verifies relative
decoherence  theories. In case of the positive detection  in
second case only it can be stated that M realizes absolute
decoherence on O and that absolute decoherence  theories are
verified.

It can be observed that suggested experiment can be very hardly
technically realized on usual measurement devices that represents
very macroscopical objects with extremely weakly manifestable
quantum characteristics. It represents basical reason why problem
of the measurement within quantum mechanics is not yet definitely
solved! But it can be pointed out that theories of the relative
decoherence does not need that M be only macroscopical. Relative
boundary can be translated toward mezoscopic, even microscopic
systems that can be conveniently chosen for measurement devices.

In further work it will be shown that Marshall {\it et al.}
experiment on the superposition of the mirror $\cite{jed}$
represents an experiment that can completely to check theories of
the absolute decoherence as well as theories of the relative
decoherence. In other words this experiment is completely able to
answer there is an absolute or relative decoherence  within
quantum mechanics and nature. In this sense Marshall {\it et al.}
experiment would be compared with remarkable Michelson experiment
that affirmed absence of the absolute space.

\section{Quantum Superposition of a Mirror
as an experimental test of the relative decoherence existence}

Suggested experimental circumstances of the quantum superposition
of a mirror is discussed in $\cite{jed}$ while theoretical basis
of this experiment is accurately given in $\cite{Ses}$-
$\cite{Dev}$. Emphasis of $\cite{jed}$, $\cite{Ses}$ -
$\cite{Dev}$ is the possibility that a (quasi)macroscopic mirror
(like Schrödinger cat $\cite{dva}$ )  be in a quantum
superposition without any absolute decoherence. Namely it is
supposed implicitly in $\cite{jed}$, $\cite{Ses}$ - $\cite{Dev}$
that absolute decoherence really exists but that it can be caused
by environmental thermal influences or by some other
unconventional way. Supposing that at extremely low temperature
(smaller than 2mK) such environmental influences can be neglected
in given mirror experiment it is concluded in $\cite{jed}$,
$\cite{Ses}$ - $\cite{Dev}$ that here absolute decoherence does
not occur. It represents a correct conclusion that, very probably,
 should be affirmed by realization of the  experiment.

But in $\cite{jed}$, $\cite{Ses}$ - $\cite{Dev}$ the possibility
that in given experiment with mirror a  relative decoherence can
appear and relative boundary principle can be affirmed is not
considered. In further work it will be shown that in the same
mirror experiment all conditions for relative decoherence are
satisfied so that it can be expected that real experiment should
to affirm relative decoherence concepts. Precisely speaking it
will be demonstrated that single photon in Michelson
interferometer can be treated as O and tiny oscillating mirror as
M in sense of the relative decoherence discussed in the previous
section of this work. In other words it will be demonstrated that
here both cases, first one with quantum superposition on O+M, and
second one with absolute or relative decoherence on O caused by M
only (since environment can be neglected) exist and can be
realistically experimentaly checked. (For relative boundary view
point environmental influences can to cause only relative
decoherence (collapse) too while quantum superposition, that
really exist on the supersystem O+M+environment is extremely
hardly experimentally checkable.)

Since in $\cite{jed}$, $\cite{Ses}$ - $\cite{Dev}$ all
experimental and theoretical details of the quantum superposition
of given mirror have been presented clearly and completely
attention of this work will be directed on the principal aspects
of given experiment only.

So, before exact quantum mechanical dynamical interaction with
tiny mirror $m$, photon $p$, that propagates through Michelson
interferometer, is in a quantum superposition
\begin {equation}
|\Psi^{p}\rangle = \frac{1}{\sqrt{2}}(|B^{p}\rangle +
|A^{p}\rangle)
\end {equation}
where $|B^{p}\rangle$ represents quantum state of $p$ propagating
through interferometer arm B without $m$ while $|A^{p}\rangle$
represents quantum state of $p$ propagating through interferometer
arm $A$ with $m$. (Given quantum superposition on $p$ can be
simply immediately detected by corresponding interference effects
detectors but such detection would represent a radical revision of
given experimental scheme.) Simultaneously $m$ is, before
interaction with $p$, described by a rested wave packet
$|0^{m}\rangle$ so that $p$+$m$ is described, before
interaction,by noncorrelated quantum state
\begin {equation}
|\Psi^{p}\rangle \otimes |0^{m}\rangle = \frac{1}{\sqrt
{2}}(|B^{p}\rangle +|A^{p}\rangle \otimes |0^{m}\rangle
\end{equation}

Exact quantum mechanical dynamical interaction between $p$ and $m$
\cite{jed}, \cite{Ses} - \cite{Dev} changes $|\Psi^{p}\rangle
\otimes |0^{m}\rangle$ during time $t$ into following quantum
state
\begin {equation}
|\Psi^{p+m}(t)\rangle = \frac{a}{\sqrt {2}} \exp (-I\omega_{p}t)
(|B^{p}\rangle \otimes |0^{m}\rangle + f(k,\omega_{m},t)
|A^{p}\rangle \otimes |g(k,\omega_{m},t)^{m}\rangle )      .
\end {equation}
Here $\omega_{p}$ represents constant $p$ frequency, $\omega_{m}$
- constant $m$ frequency, $k$ - constant parameter that quantifies
the displacement of $m$ in units of the size of the ground state
wave packet. Also, here $f(k,\omega_{m},t)= exp
(ik^{2}(\omega_{m}t- \sin \omega_{m}t))$ represents a
quasiperiodical (with period $T_{m}= \frac{2\pi}{\omega_{m}}$
function, and $|g(k, m,t)\rangle $ a periodical (with period
$T_{m}$) quantum state such that
\begin {equation}
|g(k,\omega_{m},T_{m})^{m}\rangle = |0^{m}\rangle
\end {equation}
It means that, according to (12),(13), $p$+$m$ is initially, i.e.
in zero time moment, as well as in following later  time moments
$T_{m}, 2T_{m}...$, exactly described by a noncorrelated quantum
state, while in all other time moments $p$+$m$ is exactly
described by a correlated quantum state. (More precisely speaking
there is a relatively small (in respect to $T_{m}$) time intervals
$(nT_{m}-\frac {\tau}{2},nT_{m} +\frac{\tau}{2})$ for
$n$=0,1,2,...
 whose equivalent durations $\tau$ depend from experimental
circumstances. Within any of these intervals $p$+$m$ is described
by a noncorrelated quantum state while without these intervals it
is described by a noncorrelated quantum state in a satisfactory
approximation. For this reason $\frac {\tau}{2}$ can be treated as
a time interval of an effective correlation or decorrelation.)

It is satisfied
\begin {equation}
|g^{m}\rangle \equiv |g(k,\omega_{m},t)^{m}\rangle =
|g_{-}(k,\omega_{m},t)^{m}\rangle +
|g_{+}(k,\omega_{m},t)^{m}\rangle \equiv |g_{-}^{m}\rangle +
|g_{+}^{m}\rangle
\end {equation}
where $|g_{-}^{m}\rangle \equiv |g_{-}(k,\omega_{m},t)^{m}\rangle$
and $|g_{+}^{m}\rangle \equiv |g_{+}(k,\omega_{m},t)^{m}\rangle$
are time dependent wave packets of $m$. They move periodically in
opposite directions so that distance between these two packets
becomes maximal, close to 2$k$, in time moments $\frac{T_{m}}{2},
\frac{3T_{m}}{2} ,...$ and minimal, i.e. zero in time moments 0,
$T_{m}, 2T_{m},... $ . Obviously, (14), except in time moments 0,
$T_{m}, 2T_{m},... $ (precisely time intervals $(nT_{m} -
\frac{\tau}{2},nT_{m} +\frac{\tau}{2})$  for $n$=0,1,2,... ),
represents a nontrivial quantum superposition (for this reason
$\frac{\tau}{2}$ can be treated as a time interval of the
restitution of an effective decoherence or coherence too).

Under a condition
\begin {equation}
k^{2} \geq 1
\end {equation}
momentum that $m$ obtains from $p$ is greater than initial
momentum uncertainty of $m$. Also it means that maximal distance
between two superposition terms $|g_{-}^{m}\rangle$ and
$|g_{+}^{m}\rangle$ becomes equal or greater than one size of the
ground state wave packet of $m$, $|0^{m}\rangle$. Roughly speaking
under condition (15) $m$ appears simultaneously in two places
distinct from initial. It can be added that under condition (15)
wave packets $|g_{-}^{m}\rangle$ and $|g_{+}^{m}\rangle$ become
weakly interfering not only mutually but with wave packet of the
ground state of $m$, $|0^{m}\rangle$ too.

As it has been noted, in  $\cite{jed}$ it is demonstrated that for
temperature less than 2 mK thermal influence of the environment on
$m$ can be neglected. For this reason, for given temperature,
environment cannot to generate any (absolute) decoherence on p+m.

All this admits following conclusions. Initially noncorrelated
$p$+$m$ described by (11), with $p$ in a quantum superposition
(10), evolves dynamically during time interval $(0,\frac
{\tau}{2})$ in a correlated quantum state (12). Given correlated
quantum state includes, roughly speaking, a  subsystemic effective
decoherence on $p$ and subsystemic quantum superposition
$|g^{m}\rangle$ on $m$ (precisely speaking $m$ is described by a
second kind mixture $\cite {tri}$ of initial wave packet
$|0^{m}\rangle$ and quantum superposition $|g^{m}\rangle$ (14) ).
This quantum superposition (12), without any decorrelation or
without any subsystemic coherence revival on $p$ or without any
subsystemic decoherence of (14), evolves dynamically during time
interval $(0+\frac {\tau}{2}, T_{m}-\frac {\tau}{2})$. During time
interval $(T_{m}- \frac{\tau}{2}, T_{m})$ a dynamical
decorrelation on $p$+$m$, a subsystemic coherence revival on $p$
and a subsystemic decoherence on m occur. So  in time moment
$T_{m}$ there is a complete dynamical return of $p$+$m$ in a like
initial, noncorrelated quantum state which can be simply {\it
immediately} checked by corresponding detection of the
interference effects on $p$ only. Positive detection of $p$
interference effects admits a consistent {\it mediate and
retrospective}  statement  that previously, in $(0+\frac
{\tau}{2},T_{m}-\frac {\tau}{2})$, $p$+$m$ has been in a
correlated quantum state and $m$ in a quantum superposition. But
the same positive detection represents {\it none immediate and
predictive detection} of corresponding correlated quantum state of
$p$+$m$ or corresponding susbsystemic quantum superposition of
$m$. Also, given detection representing an additional, control
measurement changes radically previous dynamical evolution on
$p$+$m$. Without this detection dynamical evolution on $p$+$m$
occurs in any later time moment periodically in an equivalent way.

On the basis of the previous analysis it can be expected that real
experiment will show that dynamical interaction between $p$ and
$m$ does not cause any absolute decoherence on $p$+$m$ in any time
moment.

But it is not hard to see that during
$(0+\frac{\tau}{2},T_{m}-\frac {\tau}{2})$, when $p$+$m$ is in
correlated quantum state (12), all conditions for an effective
approximation of this state by relatively decoherent state
(presented in previous section of this work) are satisfied.

{\it Firstly}, $p$+$m$ is really described by a correlated quantum
state $|\Psi^{p+m}(t)\rangle$ (12).

{\it Secondly}, $m$ is effectively subsystemically described by a
second kind mixture of the initial wave packet $|0^{m}\rangle$ and
$|g^{m}\rangle$ representing an quantum superposition of the wave
packets $|g_{-}^{m}\rangle$ and $|g_{+}^{m}\rangle$ .

{\it Thirdly }, according to (12) there is one-to-one correlation
between $|B^{p}\rangle$ and $|0^{m}\rangle$  from one, and between
$|A^{p}\rangle$ and $|g^{m}\rangle$ from other side.

{\it Fourth}, according to (15), all three wave packets
$|0^{m}\rangle$,$|g_{-}^{m}\rangle$ and $|g_{+}^{m}\rangle$ are
mutually weakly interfering.

For this reason, according to theorem of the selfdecoherence,
during $(0+\frac {\tau}{2},T_{m}+\frac{\tau}{2})$ on
subsystemically effectively approximately, i.e. "classicaly"
described M effective approximate selfdecoherence exists.
Simultaneously, according to quantum correlation (12), in relation
to selfdecoherent  $m$ on subsystemically effectively quantum
mechanically exactly described $p$ effective exact  relative
decoherence exists. This  effective and relative decoherence on
$p$ can be certainly affirmed by additional control measurement,
i.e. by detection of the complete absence of the interference
effects on $p$ only during given time interval. Certainty of the
detection of absence of interference effects on $p$ can be, in
sense of Einstein criterion $\cite{Des}$, used as a sufficient
proof that before detection $p$ is effectively in a relatively
decoherent state caused by selfdecoherence on $m$. Thus, in sense
of Einstein criterion given detection is neither mediate nor
retrospective, but of course such detection as an additional
measurement changes radically previous dynamical evolution on
$p$+$m$. Of course, equivalent detection in time moment $T_{m}$,
in sense of previous discussion, can to affirm that previous
decoherence is really effective and relative but not absolute.

All this admits that during $(0+\frac {\tau}{2},T_{m}-\frac
{\tau}{2})$ exact correlated quantum state $|\Psi^{p+m}(t)\rangle$
(12) on $p$+$m$ can be effectively approximated by a discretely
different hybrid description of $p$+$m$ corresponding to a
relative collapse on $p$ as O caused by selfdecoherent $m$ as M.
In this way $m$ is not only a (quasi)macroscopic system in a
quantum superposition but it is, complementary, a
(quasi)microscopic measurement device. Also, it can be pointed out
that here time interval $(0,\frac {\tau}{2})$ can be treated in
two discretely different ways. Exactly quantum mechanically
dynamically it represents  the time interval of the dynamical
realization of the correlation between $p$ and $m$. But
effectively approximately quantum mechanically, or "classically"
it represents a time interval of the relative decoherence
realization.

In this way it is proved that Marshall {\it et al.} experiment of
the quantum superposition of a mirror represents an experimental
test that should to affirm the relative decoherence concept, i.e.
relative boundary principle and completeness of the standard
quantum mechanical formalism.

\section{Conclusion}
In conclusion following can be only repeated and pointed out.
Marshall {\it et al.} experiment of the quantum superposition of a
mirror should to testify not only a (quasi)macroscopic quantum
superposition. It should to testify also relative decoherence (as
a spontaneous unitary symmetry (superposition) breaking (effective
hiding) ), i.e. relative boundary principle and completeness of
the standard quantum mechanical formalism. In this sense it would
have same significance in foundation of the quantum mechanics as
well as remarkable Michelson experiment in foundation of the
special theory of relativity. It is a curiosity that both
experiments use practically identical experimental circumstances
and that they negate absolute concepts, i.e. concepts of the
fundamental symmetries absolute breaking.

\vspace {1 cm}

{\bf  Acknowledgements}
\vspace {1 cm}

Authors are deeply grateful to Prof.Dr. Fedor Herbut whose
personal views on the relative decoherence (collapse) has been
very instructive and inspirative for this work and to  Prof.Dr.
Darko Kapor and Prof.Dr. Milan Vuji\v {c}i\'c on the
illuminating discussions,support and help.

\vspace {1 cm }

\end{document}